\documentclass{pasj00}
\begin{document}
\SetRunningHead{H.~Yamaguchi, A.~Bamba, and K.~Koyama}{The largest non-thermal SNR 30~Dor~C with Suzaku}
\Received{2008/06/30}
\Accepted{2008/07/26}
\title{Suzaku Observation of 30~Dor~C: 
A Supernova Remnant with the Largest Non-Thermal Shell}
\author{Hiroya \textsc{Yamaguchi}}
\affil{RIKEN (The Institute of Physical and Chemical Research), 
  2-1 Hirosawa, Wako, Saitama 351-0198}
\email{hiroya@crab.riken.jp}
\author{Aya \textsc{Bamba}}
\affil{Institute of Space and Astronautical Science, JAXA, 
  3-1-1 Yoshinodai, Sagamihara, Kanagawa 229-8510}
\author{and \\
Katsuji \textsc{Koyama}}
\affil{Department of Physics, Kyoto University, 
  Kitashirakawa-oiwake-cho, Sakyo-ku, Kyoto 606-8502}
\KeyWords{ISM:~individual~(30~Dor~C) --- supernova remnants --- X-Rays:~spectra} 
\maketitle

\begin{abstract}

This paper reports on the Suzaku results of thermal and non-thermal 
features of 30~Dor~C, a supernova remnant (SNR) in a superbubble of 
the Large Magellanic Cloud (LMC). 
The west rim exhibits a non-thermal X-ray spectrum with no thermal 
component. A single power-law model is rejected but a power-law model 
with spectral cutoff is accepted. The cutoff frequency of 
$(3-7)\times 10^{17}$~Hz is the highest among the shell type SNRs 
like SN 1006 ($\sim 6\times 10^{16}$~Hz), and hence 30~Dor~C would be 
the site of the highest energy accelerator of the SNR shock. 
The southeast (SE) and northeast (NE) rims have both the thermal and 
non-thermal components. The thin-thermal plasmas in the both rims 
are in collisional ionization equilibrium state. The electron 
temperature of the plasma in the SE rim ($kT_e \sim 0.7$~keV) is 
found to be higher than the previously reported value. 
The power-law index from SE is nearly the same as, 
while that from the NE is larger than that of the West rim.
The SNR age would be in the range of $(4-20)\times 10^3$~yr. 
Thus, 30~Dor~C is likely to be the oldest shell-like SNR with 
non-thermal emission. 

\end{abstract}

\section{Introduction}
\label{sec:introduction}

OB associations typically contain a few 10 massive stars. The combined 
actions of fast stellar winds and core-collapse supernova (SN) explosions 
of the member stars create large ($\gtrsim 10$~pc) shell-like structures, 
called superbubbles (SBs), by sweeping up the ambient medium 
(e.g., Mac Low \& McCray 1988). Therefore, SBs are very energetic objects 
which store a large amount of energy injected from massive stars; 
the swept up interstellar medium (ISM) can be successively heated to 
high temperature by  stellar winds and/or SNe.
Furthermore, large tenuous cavities created inside the SB walls allow 
that the blast shocks of the interior supernova remnants (SNRs) expand 
rapidly without decelerating for a long time. Therefore, the timescale 
of efficient cosmic-ray acceleration can be much longer than that of
most isolated SNRs.

For the study of SNRs and SBs, the Large Magellanic Cloud (LMC) is 
an ideal site, because of little foreground extinction and well-known 
distance (50~kpc: Alves 2004). Dunne, Points, Chu~(2001) systematically 
studied several SBs in the LMC using ROSAT, and showed their X-ray 
emissions are brighter than that theoretically expected for a wind-blown 
bubble. This fact suggests that the X-ray emitting walls of the SBs are 
re-heated by the shock of interior SNRs, and hence the emissions are 
enhanced. They also found that the plasma temperatures of the shocked 
SB walls are typically of the order of 0.1~keV.

30~Dor~C, located in the southwestern region of the 30~Doradus complex, 
is one of the SBs in the LMC identified by the morphology of the 
H$\alpha$ emission (Mathewson et al.~1985; Kennicutt \& Hodge 1986). 
This SB hosts an OB association LH~90, with the age of a few Myr 
(Lucke \& Hodge 1970). A shell-like structure with a diameter of 
$\sim \timeform{6'}$ ($\sim 87$~pc at 50~kpc distance) was 
discovered by the radio observation (Mills et al.~1984).

X-rays from 30~Dor~C were first detected with the 
Einstein satellite (Long et al.~1981). Then, 
Dunne, Points, Chu~(2001) found a clear shell-like structure 
confined within the H$\alpha$ shell by the ROSAT observation. 
They thus concluded that the X-ray emission arises from the 
interaction between the SB and an interior SNR. 
Bamba et al.~(2004: hereafter B04) discovered synchrotron X-rays from 
the shell region of 30~Dor~C with Chandra and XMM-Newton. They found 
that the total luminosity of the synchrotron emission was $\sim 10$ 
times larger than that of SN~1006 (Koyama et al.~1995), and hence argued 
that the large energy supply was probably by multiple SN explosions. 


This paper reports the X-ray observation of 30~Dor~C with Suzaku 
(Mitsuda et al.~2007).  Utilizing the good sensitivity of Suzaku, 
we study the detailed properties of the thermal and non-thermal 
emissions from 30~Dor~C. The distance to the LMC is 
assumed to be 50~kpc, following Alves~(2004).

\section{Observations and Data Reduction}
\label{sec:observation}

Suzaku observed SN~1987A on 2005 November 3 (Observation ID = 500006010; 
hereafter Obs.~1) and  on 2006 June 8 (Observation ID = 801090010; 
hereafter Obs.~2). In both of the observations, the 30~Dor~C region was 
in the field of view (FOV) of X-ray Imaging Spectrometer 
(XIS: Koyama et al.~2007). This paper concentrates on the imaging and 
spectral features of 30~Dor~C obtained with the XIS. 
The results of SN~1987A will be reported in separate paper.

Three of the XIS are front-illuminated (FI) CCDs and the remaining 
one is a back-illuminated (BI) CCD. The former have high sensitivity 
and low background for diffuse sources in the hard X-ray band of 
$\sim$~5--10~keV, while the latter has high sensitivity in the soft 
X-ray band below 1~keV. All the CCDs are placed 
on the foci of the X-Ray Telescopes (XRT: Serlemitsos et al.~2007) which 
co-aligned to image the same region of the sky. A half-power diameter (HPD) 
of the XRT is $\sim \timeform{1.8'}-\timeform{2.3'}$, independent of 
the X-ray energy.

The XIS was operated in the normal full-frame clocking mode during both of 
the observations. We employed cleaned revision 2.0 data, which includes 
column-to-column correction of the charge transfer inefficiency (CTI) 
measured using the charge-injection (CI) capability 
(Nakajima et al.~2008). We used the HEADAS software 
version 6.4 for the data reduction and analysis. After the screening, 
the exposure times were obtained to be $\sim$~37~ks and $\sim$~40~ks for 
Obs.~1 and Obs.~2, respectively.

The errors quoted in the text and tables are at the 90\% confidence level, 
and the 1$\sigma$ confidence in the figures, unless otherwise stated.

\section{XIS Image}
\label{sec:image}

\begin{figure}[t]
  \begin{center}
    \FigureFile(80mm,80mm){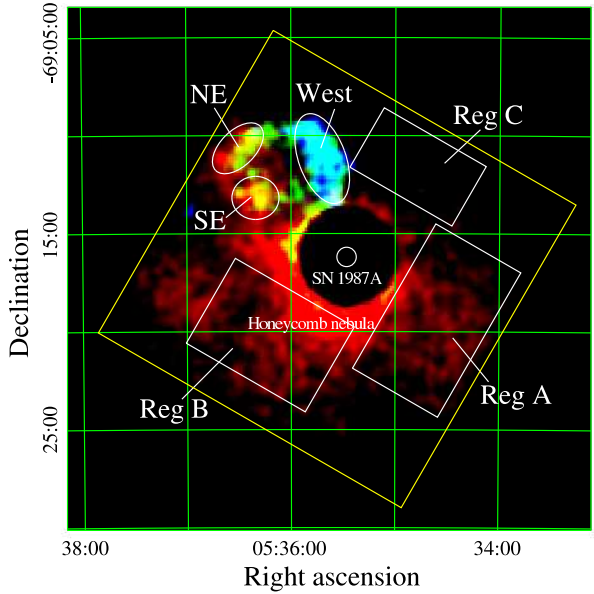}
  \end{center}
  \caption{Three-color XIS image around SN~1987A and 30~Dor~C 
    smoothed with a Gaussian kernel of $\sigma$ = \timeform{25''}. 
    Red, green, and blue contain emissions from 0.6--1.0~keV, 
    1.0--2.0~keV, and 2.0--5.0~keV, respectively. 
    The coordinates (RA and Dec) refer to epoch J2000.0. 
    The Field of view of the XIS is indicated by the yellow square, 
    where the data from the four XIS are combined, 
    but that of the four corners, irradiated by 
    the $^{55}$Fe calibration sources, are removed. 
    The ellipses and rectangles indicate the regions used in 
    our spectral analysis (section~\ref{sec:spectrum}). 
    The small circle represents the position of SN~1987A, 
    but the \timeform{2.5'}-radius circular region around 
    SN~1987A is excluded (see text). 
  }  \label{fig:image}
\end{figure}

Figure~\ref{fig:image} shows the three-color XIS image: red, green, 
and blue represent 0.6--1.0~keV, 1.0--2.0~keV, and 2.0--5.0~keV, 
respectively. The XIS astrometry was fine-tuned using SN~1987A at 
(RA, Dec) = (\timeform{05h35m28s},\timeform{-69D16'12"}). 
After the tuning, we excluded the circular region with the radius of 
\timeform{2.5'} around SN~1987A to emphasize dim diffuse structures. 

A clear shell-like structure of 30~Dor~C can be seen at the northeast 
part of the FOV. The west side of the shell is bright in the hard X-ray 
band, whereas the eastern half is dominated by the soft X-rays bellow 2 keV. 

In addition, we can see extended soft X-ray emission from the entire FOV, 
with different surface brightness from position to position. 

\section{Spectral Analysis}
\label{sec:spectrum}

For a spatially resolved X-ray spectroscopy of 30~Dor~C, we selected
typical source regions and candidates of a background region.
To increase statistics, the data obtained in Obs.~1 and Obs.~2 
were merged both for the source and background spectra.

The selected 30~Dor~C regions are shown with the three ellipses in 
figure~\ref{fig:image}, the West, southeast (SE), and northeast (NE) 
regions. West was chosen as the brightest hard X-ray (2.0--5.0~keV) 
region, while the other two were selected from the peaks of the soft 
X-ray band (0.6--1.0~keV).  

\subsection{Background Selection and Analysis Procedure}
\label{ssec:method}

\begin{figure}[t]
  \begin{center}
    \FigureFile(80mm,80mm){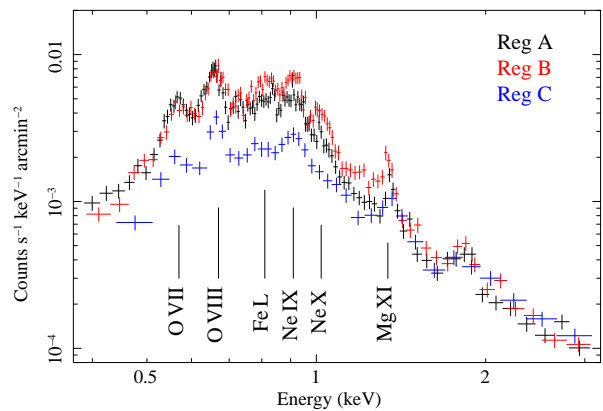}
  \end{center}
  \caption{XIS-BI spectra of Regions~A (black), B (red) and C (blue), 
    normalized by each solid angle. 
    Characteristic emission lines are labeled in the panel. 
  }
  \label{fig:compare}
\end{figure}

For a proper background subtraction, we examined the spectra of 
source-free regions shown by rectangles (Region~A, B, and C) in 
figure~\ref{fig:image}. Figure~\ref{fig:compare} shows the XIS-BI spectra 
of these regions normalized by each solid angle. 
Although the spectral shapes above $\sim 1.5$~keV are very similar among 
the three rectangular regions (Region~A, B, and C), the 0.4--1.0~keV 
surface brightness of Region~C is lower, while the 0.7--1.5 keV flux 
of Region~B is higher than those of the others.

Since a dense molecular cloud is located at the north region of our FOV 
(the NANTEN CO observation: Yamaguchi et al.~2001), 
the low energy (0.4--1.0 keV) deficiency in Region~C would 
be due to the local interstellar absorption in the LMC.

The excess in 0.7--1.5 keV of Region~B spectrum is particularly 
clear at the energies of the K-shell emission lines of 
Ne\emissiontype{IX}, Ne\emissiontype{X}, and Mg\emissiontype{XI}, 
and Fe L-shell blends. These features are likely to originate from 
thermal emissions from the Honeycomb nebula (SNR~0536--69.3: 
Dennerl et al.~2001), which is partially included in the Region~B area, 
although no clear structure is found in figure~\ref{fig:image}.

We thus assumed that Region~A represents a proper local background 
for 30~Dor~C. However, the local absorption may be variable from position 
to position, and hence we applied the spectra analysis as follows.

(1) Non X-ray background (NXB) was constructed from the data base
of night-Earth by sorting with the geomagnetic cut-off rigidity (COR) from 
the same region on the detector coordinates of our observation 
(see Tawa et al.~2008). The NXB with the same COR distribution as 
that of the 30~Dor~C observations were subtracted from the spectra of 
Region~A, and the West, SE, and NE regions. 
 
(2) The NXB-subtracted spectrum of Region~A was fitted with a model of 
the cosmic X-ray background (CXB) plus several thermal plasma components. 
The best-fit spectrum was used as a local background for the source 
spectra (West, SE, and NE). 

(3) Since the source spectra include a local background, we performed
model fit, by adding the Region~A components and fixing the parameters 
other than absorptions to those of the best-fit results obtained in 
the process (2).

For the spectral fittings, we used the XSPEC software version 11.3.2. 
The data from the three XIS-FIs were merged to improve the photon 
statistics, since their responses are nearly identical. 
For simplicity, the XIS-BI spectra are not shown in figures, 
although they were analyzed simultaneously with the FI spectra.

\subsection{Background Region (Region A)}
\label{ssec:bgd}

\begin{figure}[t]
  \begin{center}
    \FigureFile(80mm,80mm){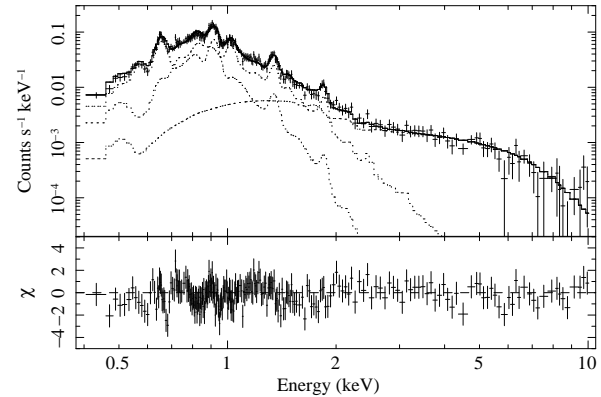}
  \end{center}
  \caption{XIS-FI spectrum of Region~A. The non X-ray background was 
    subtracted. The components of the best-fit model are shown with 
    the dotted lines. 
  }
  \label{fig:bgd}
\end{figure}

\begin{table}[t]
  \begin{center}
  \caption{Best-fit spectral parameters of Region~A.}
    \label{tab:bgd}
    \begin{tabular}{llc}
      \hline
  Component  &  Parameter & Value  \\ 
  \hline
  Absorption$^{\ast}$  & $N_{\rm H}^{\rm LMC}$ (cm$^{-2}$) & 
                         5.3 (0.9--9.6) $\times 10^{20}$  \\
  Power-law$^{\dagger}$ & $\Gamma$         & 1.412 (fixed)  \\
  ~                    & Norm$^{\ddagger}$ & 3.7 (3.5--4.0) $\times 10^{-5}$ \\
  VAPEC   & $kT_{e1}$ (keV)$^{\S}$ & 0.22 (0.21--0.24)  \\
  ~       & EM$_1$ (cm$^{-3}$)$^{\S\#}$  & 5.9 (3.8--8.7) $\times 10^{58}$  \\
  ~       & $kT_{e2}$ (keV)$^{\|}$  & 0.53 (0.48--0.56)  \\
  ~       & EM$_2$ (cm$^{-3}$)$^{\|\#}$ & 2.0 (1.5--2.6) $\times 10^{58}$  \\
  ~       & O  (solar) & 0.30 (0.23--0.41)  \\
  ~       & Ne (solar) & 0.61 (0.47--0.77)  \\
  ~       & Mg (solar) & 0.53 (0.41--0.72)  \\
  ~       & Si (solar) & 0.50 (0.37--0.66)  \\
  ~       & Fe (solar) & 0.23 (0.19--0.30)  \\
  \hline
  $\chi ^2$/d.o.f.  & ~  & 427/417 = 1.02   \\
  \hline 
  \multicolumn{3}{l}{\parbox{80mm}{\footnotesize
      $^{\ast}$ Absorption in the LMC (see text).}} \\
  \multicolumn{3}{l}{\parbox{80mm}{\footnotesize
    $^{\dagger}$ Cosmic X-ray background is assumed.}} \\
  \multicolumn{3}{l}{\parbox{80mm}{\footnotesize
    $^{\ddagger}$ Unit is photons~s$^{-1}$~cm$^{-2}$~keV$^{-1}$ at 1~keV.}} \\ 
  \multicolumn{3}{l}{\parbox{80mm}{\footnotesize
      $^{\S}$ Low temperature component.}} \\
  \multicolumn{3}{l}{\parbox{80mm}{\footnotesize
      $^{\|}$ High temperature component.}} \\
  \multicolumn{3}{l}{\parbox{80mm}{\footnotesize
      $^{\#}$ Emission measure (EM = $n_p n_e V$), where $n_p$, $n_e$, 
      and $V$ are the proton and electron densities and 
      the plasma volume, respectively.}} \\
  \end{tabular}
 \end{center}
\end{table}

Figure~\ref{fig:bgd} shows the 0.4--10~keV spectrum of Region~A. 
We can see many emission lines below $\sim 2$~keV, 
while the spectrum above $\sim 2$~keV is featureless. 
Therefore, we fitted the spectrum with a model of a thin-thermal plasma 
plus a power-law component. For the thermal plasma, we used a VAPEC 
model (Smith et al.~2001), allowing the electron temperature ($kT_e$) 
and emission measure (EM) to vary freely. The elemental abundances 
relative to solar values (Anders \& Grevesse 1989) of O, Ne, Mg, Si, 
and Fe were also treated as free parameters, whereas those of the other 
elements were fixed to the mean LMC values of Russell \& Dopita (1992).

Since the power-law component is mainly due to the CXB, we fixed the photon 
index to be $\Gamma$ = 1.412, which was reported by Kushino et al.~(2002). 
The interstellar absorption columns in the Galaxy and LMC were separately 
treated. The Galactic absorption column density was fixed to be 
$N_{\rm H}^{\rm G} = 6.35\times 10^{20}$~cm$^{-2}$ (Dickey \& Lockman 1990), 
assuming the solar abundances. 
[Although the Galactic H\emissiontype{I} column densities were newly 
measured by Leiden/Argentine/Bonn survey (Kalberla et al.~2005), their 
database does not exclude the contribution of H\emissiontype{I} in the 
LMC.\footnote{see $<$http://heasarc.gsfc.nasa.gov/Tools/w3nh\_help.html$>$, 
for details.} Therefore, we did not use this new database.] 
The absorption in the LMC ($N_{\rm H}^{\rm LMC}$) with the LMC abundances 
(Russell \& Dopita 1992) was treated as a free parameter.

This 2-component model gave the best-fit $kT_e$ of 0.29~keV, while a 
significant data excess at the energy of the O\emissiontype{VII} K$\alpha$ 
line ($\sim 0.57$~keV) are found. In fact, this model was rejected with 
$\chi ^2$/d.o.f.~of 628/419. 
We hence added another thermal component with a different temperature. 
The metal abundances were common for both components. Then, the fit 
was greatly improved with $\chi ^2$/d.o.f.~of 427/417. The best-fit 
parameters and models are respectively shown in table~\ref{tab:bgd}, and 
figure~\ref{fig:bgd} with dotted lines. The 2--10~keV surface brightness 
of the power-law component was obtained to be 
$6.6\times 10^{-8}$~erg~cm$^{-2}$~s$^{-1}$~sr$^{-1}$, 
which is consistent with being the CXB (Kushino et al.~2002).

\subsection{West Rim}
\label{ssec:west}

\begin{figure}[t]
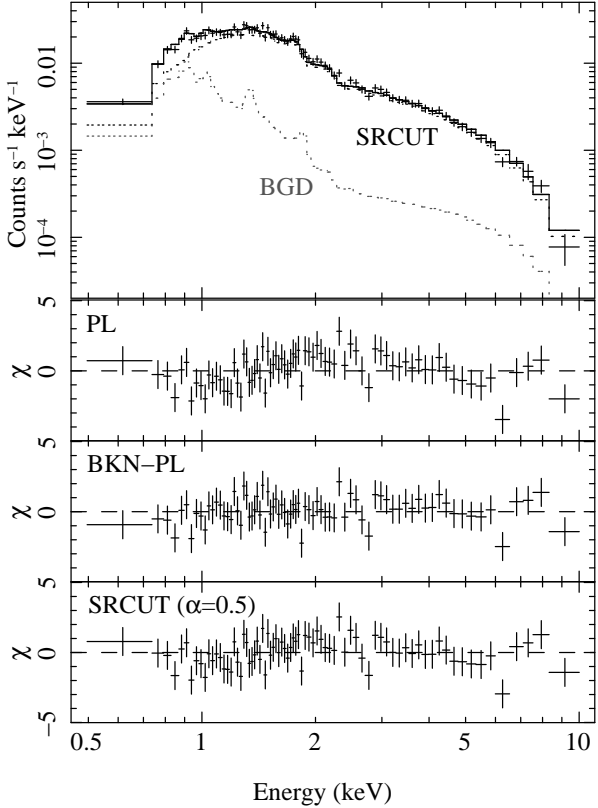

  \begin{center}
    \FigureFile(80mm,80mm){Figure/figure4.ps}
  \end{center}
  \caption{Non X-ray background subtracted XIS-FI spectrum of 
    the West rim. In the top panel, the best-fit SRCUT model is 
    shown with the black dotted line, while the local background 
    component is represented by the gray dotted line. 
    The second, third, and forth panels display the residuals 
    from the best-fits of the power-law (PL), broken power-law 
    (BKN-PL), and SRCUT ($\alpha = 0.5$) models, respectively. 
  }
  \label{fig:west}
\end{figure}

\begin{table}[t]
  \begin{center}
  \caption{Best-fit parameters for the West spectrum.}
    \label{tab:west}
    \begin{tabular}{lc}
      \hline 
  Parameter &  Value  \\
  \hline
  \multicolumn{2}{c}{Power-law (PL)} \\
  \hline
  $N_{\rm H}^{\rm LMC}$ (cm$^{-2}$)$^{\ast}$ & 5.9 (5.6--6.2) $\times 10^{21}$  \\
  $\Gamma$  & 2.17 (2.12--2.21) \\
  Norm$^{\dagger}$  & 2.8 (2.7--2.9) $\times 10^{-4}$  \\
  $F_{\rm X}$ (ergs~s$^{-1}$~cm$^{-2}$)$^{\ddagger}$  & $1.2\times 10^{-12}$ \\
  $\chi ^2$/d.o.f.  & 191/148 = 1.29 \\
  \hline
  \multicolumn{2}{c}{Broken power-law (BKN-PL)} \\
  \hline
  $N_{\rm H}^{\rm LMC}$ (cm$^{-2}$)$^{\ast}$ & 4.4 (4.2--5.1) $\times 10^{21}$  \\
  $\Gamma _1$  & 1.48 (1.31--1.82) \\
  Break $E$ (keV) & 1.9 (1.7--2.5) \\
  $\Gamma _2$  & 2.33 (2.26--2.50) \\
  Norm$^{\dagger}$  & 2.0 (1.9--2.3) $\times 10^{-4}$ \\
  $F_{\rm X}$ (ergs~s$^{-1}$~cm$^{-2}$)$^{\ddagger}$  & $9.9\times 10^{-13}$ \\
  $\chi ^2$/d.o.f.  & 140/146 = 0.96 \\
  \hline
  \multicolumn{2}{c}{SRCUT ($\alpha$ = 0.50)} \\
  \hline
  $N_{\rm H}^{\rm LMC}$ (cm$^{-2}$)$^{\ast}$ & 5.5 (5.2--5.7) $\times 10^{21}$  \\
  $\nu _{\rm rolloff}$ (Hz)  & 3.7 (3.2--4.5) $\times 10^{17}$  \\
  Norm$^{\S}$  & 8.0 (7.0--9.0) $\times 10^{-3}$  \\
  $\chi ^2$/d.o.f.  & 164/148 = 1.11 \\
  \hline
  \multicolumn{2}{c}{SRCUT ($\alpha$ = 0.55)} \\
  \hline
  $N_{\rm H}^{\rm LMC}$ (cm$^{-2}$)$^{\ast}$ & 5.5 (5.2--5.8) $\times 10^{21}$  \\
  $\nu _{\rm rolloff}$ (Hz)  & 4.6 (3.8--5.6) $\times 10^{17}$  \\
  Norm$^{\S}$  & 1.9 (1.7--2.1) $\times 10^{-2}$  \\
  $\chi ^2$/d.o.f.  & 166/148 = 1.12 \\
  \hline
  \multicolumn{2}{c}{SRCUT ($\alpha$ = 0.60)} \\
  \hline
  $N_{\rm H}^{\rm LMC}$ (cm$^{-2}$)$^{\ast}$ & 5.5 (5.3--5.8) $\times 10^{21}$  \\
  $\nu _{\rm rolloff}$ (Hz)  & 5.7 (4.7--7.1) $\times 10^{17}$  \\
  Norm$^{\S}$  & 4.6 (4.1--5.2) $\times 10^{-2}$  \\
  $\chi ^2$/d.o.f.  & 167/148 = 1.13 \\
  \hline 
  \multicolumn{2}{l}{\parbox{65mm}{\footnotesize
    $^{\ast}$ Absorption in the LMC.}} \\
  \multicolumn{2}{l}{\parbox{65mm}{\footnotesize
    $^{\dagger}$ Unit is photons~s$^{-1}$~cm$^{-2}$~keV$^{-1}$ at 1~keV.}} \\ 
  \multicolumn{2}{l}{\parbox{65mm}{\footnotesize
    $^{\ddagger}$ Unabsorbed flux in 0.5--10~keV band.}} \\
  \multicolumn{2}{l}{\parbox{65mm}{\footnotesize
    $^{\S}$ Flux density at 1~GHz (Jy).}} \\
  \end{tabular}
 \end{center}
\end{table}

Figure~\ref{fig:west} shows the spectrum of the 30~Dor~C West region. 
Since the spectrum is featureless with no emission line, we first fitted 
with a power-law (PL) model plus the local background derived in 
subsection~\ref{ssec:bgd}.  Then, the obtained $\chi ^2$/d.o.f. was 
191/148, hence a simple PL model is rejected. It shows the residuals 
with wavy shape around the best-fit power-law of photon index 
$\Gamma \sim 2.2$, as is shown in the second panel of figure~\ref{fig:west}. 

We therefore, introduced a broken power-law (BKN-PL) model, Then, the 
$\chi ^2$/d.o.f.~was significantly reduced to 140/146 to become an acceptable 
fit. The residual is shown in the third panel of figure~\ref{fig:west}; 
the wavy pattern is largely reduced. The photon indices of the BKN-PL model 
were respectively  $\Gamma _1 \sim 1.5$ and $\Gamma _2 \sim 2.3$, 
below and above a break energy at $\sim 1.9$~keV. 

We also fitted the spectrum with a more physical model, the SRCUT model 
(Reynolds 1998), which represents the synchrotron spectrum from electrons 
with an exponentially cut-off power-law energy distribution. 
We fixed the spectral index at 1~GHz to a typical value for SNRs of 
$\alpha = 0.50$. This model was also acceptable with 
$\chi ^2$/d.o.f.~= 164/148, although the residual pattern 
(the fourth panel of figure~\ref{fig:west}) was not removed completely.  
We further tried the SRCUT model for the radio indices of $\alpha = 0.55$ 
and 0.60. Both of the cases gave almost the same result as the $\alpha = 0.5$ 
model. The best-fit parameters of each model which we used here are given 
in table~\ref{tab:west}.

\subsection{SE Rim}
\label{ssec:se}

\begin{figure}[t]
  \begin{center}
    \FigureFile(80mm,80mm){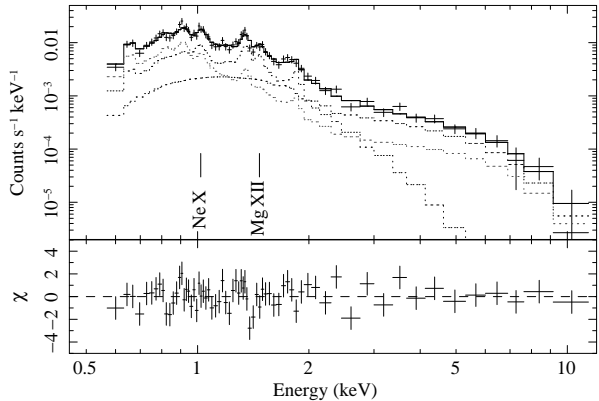}
  \end{center}
  \caption{Same as figure~\ref{fig:west}, but for the SE rim. 
    The components of the best-fit model 
    are shown with the black dotted lines. The gray dotted line 
    represents the local background component. 
    The energies of the emission lines from Ne\emissiontype{X} 
    and Mg\emissiontype{XII} are indicated in the panel. 
  }
  \label{fig:se}
\end{figure}

The spectrum of the SE region is shown in figure~\ref{fig:se}. 
We can see the relatively strong emission lines below $\sim 2.0$~keV, 
which suggests that the soft X-rays dominantly comes from an optically 
thin-thermal plasma. Indeed, B04 showed that the Chandra and XMM-Newton 
spectra of this region were well-reproduced by the model of the 
thin-thermal plasma at a collisional ionization equilibrium (CIE) plus 
the power-law component. Therefore, we fitted our spectrum using a VAPEC 
model (for the soft component), a power-law (for the hard component), 
and the local background. We allowed the abundances of 
Ne, Mg, Si, and Fe to be free. The abundance of Ni was fixed to 
that of Fe. Then, the fit was acceptable with  
$\chi ^2$/d.o.f.~of 102/101. The best-fit parameters and models are 
shown in table~\ref{tab:se_ne} and figure~\ref{fig:se}, respectively. 

Since the error region of the photon index ($\Gamma$) includes the 
best-fit value of the West rim, we fitted with the same model but 
fixing $\Gamma$ to be 2.17, then the best-fit result were almost the 
same as the case of free parameter of $\Gamma$. Although we tried to 
fit the spectrum using a non-equilibrium ionization (NEI) plasma model 
instead of the VAPEC model, the fit did not improve significantly.

\begin{table*}[t]
  \caption{Best-fit parameters of for the SE and NE spectra.}
  \begin{center}
    \label{tab:se_ne}
    \begin{tabular}{llcc}
      \hline 
  Component  &  Parameter &   SE   &  NE   \\
  \hline
  Absorption$^{\ast}$ & $N_{\rm H}^{\rm LMC}$ (cm$^{-2}$) & 
      6.2 (2.8--9.7) $\times 10^{20}$ &  3.1 (2.4--4.8) $\times 10^{21}$ \\
  Power-law & $\Gamma$ & 2.01 (1.39--2.45)  & 2.56 (2.45--2.71)  \\
  ~         & Norm$^{\dagger}$   & 2.0 (0.7--3.8) $\times 10^{-5}$  & 
                         1.2 (1.1--1.5) $\times 10^{-4}$      \\
  ~         & $F_{\rm X}$ (ergs~s$^{-1}$~cm$^{-2}$)$^{\ddagger}$  & 
                        $9.3\times 10^{-14}$ & $4.2\times 10^{-13}$ \\
  VAPEC   & $kT_e$ (keV)  & 0.66 (0.58--0.76) & 0.28 (0.20--0.34) \\
  ~       & EM (cm$^{-3}$)$^{\S}$ & 7.8 (3.2--12) $\times 10^{58}$ &  
                               9.2 (3.7--31) $\times 10^{57}$   \\
  ~       & Ne (solar) & 0.33 (0.17--0.88) & 0.53 (0.30--1.1) \\
  ~       & Mg (solar) & 0.63 (0.38--1.6)  & 1.1  (0.43--4.0) \\
  ~       & Si (solar) & 0.36 (0.19--0.96) & 3.0  ($>$0.70)   \\
  ~       & Fe, Ni (solar) & 0.03 (0.01-0.08)  & ($<$0.2)    \\
  \hline
  $\chi ^2$/d.o.f.  & ~ & 102/101 = 1.01 & 120/116 = 1.04   \\
  \hline 
  \multicolumn{4}{l}{\parbox{120mm}{\footnotesize
    $^{\ast}$ Absorption in the LMC.}} \\
  \multicolumn{4}{l}{\parbox{120mm}{\footnotesize
    $^{\dagger}$ Unit is photons~s$^{-1}$~cm$^{-2}$~keV$^{-1}$ at 1~keV.}} \\ 
  \multicolumn{4}{l}{\parbox{120mm}{\footnotesize
    $^{\ddagger}$ Unabsorbed flux in 0.5--10~keV band.}} \\
  \multicolumn{4}{l}{\parbox{120mm}{\footnotesize
    $^{\S}$ Emission measure (EM = $n_p n_e V$), where $n_p$, $n_e$, and $V$ are 
    the proton and electron densities and the plasma volume, respectively.}} \\
  \end{tabular}
 \end{center}
\end{table*}

\subsection{NE Rim}
\label{ssec:ne}

\begin{figure}[t]
  \begin{center}
    \FigureFile(80mm,80mm){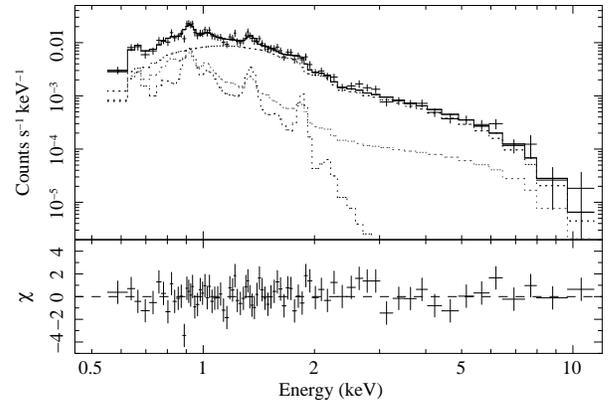}
  \end{center}
  \caption{Same as figure~\ref{fig:se}, but for the NE rim.}
  \label{fig:ne}
\end{figure}

The spectrum of NE rim is shown in figure~\ref{fig:ne}. 
Since B04 reported that no significant thermal component was 
detected from this region, we first fitted the spectrum with a 
power-law and the local background. However, this model was 
rejected with a $\chi ^2$/d.o.f.~of 166/122. The large residuals 
were found at the energy of the Ne\emissiontype{IX} K$\alpha$ line 
($\sim 0.91$~keV). This fact suggests the presence of a thermal 
component in the soft X-ray band. Therefore, we added a VAPEC model on 
the power-law. The free parameters of the VAPEC component were the same 
as described in subsection~\ref{ssec:se}. 
Then the fit was significantly improved with $\chi ^2$/d.o.f.~of 120/116. 
The best-fit parameters and models are respectively given in 
table~\ref{tab:se_ne} and figure~\ref{fig:ne}.

For the NE rim, if we fitted with the same model of fixed $\Gamma$ = 2.17 
as the case of the SE rim, then the fit was rejected with 
$\chi^ 2$/d.o.f.~of 142/117. Thus, unlike the SE rim, the NE rim has 
a steeper (softer) photon index than the West. We also tried an NEI model, 
but no improvement of $\chi ^2$/d.o.f.~value was obtained.

\section{Discussion}
\label{sec:discussion}

\subsection{Thermal Emission}
\label{ssec:thermal}

The thermal component of the SE spectrum was well represented by the 
CIE plasma model with the electron temperature of $\sim 0.7$~keV. 
Since the bright H$\alpha$ emission coincides with the SE rim 
(Mathewson et al.~1985), the origin of the thermal component is 
considered to be the SB shell re-shocked by the blast-wave of 
the interior SNR. In addition, we detected the significant 
thin-thermal component from the NE rim, for the first time. 
This region is also correlated with the H$\alpha$ emission, 
albeit the brightness is lower than that of the SE region. 
Therefore, the same origin with the SE can be considered. 
The low elemental abundance of Fe relatively to the other lighter 
elements is probably due to that the SB shell is dominantly 
contributed by stellar winds and core-collapse SNRs occurred 
in the past.

The size of the SE elliptic region is 
$\timeform{1.2'} \times \timeform{1.1'}$, 
which corresponds to 18~pc $\times$ 16~pc at a distance of 50~kpc.
Assuming the plasma depth to be 22~pc, a half radius of the SNR, 
the emission volume is estimated to be 
$V = 5.7 \times 10^{59}$~cm$^3$. 
Thus, the EM of the VAPEC component in the SE region corresponds 
to the proton and electron densities of 
$n_p \simeq n_e = 0.37~f_{\rm SE}^{-0.5}$~cm$^{-3}$, 
where $f_{\rm SE}$ is the filling factor. The plasma mass is, 
therefore, estimated to be $M = 175~f_{\rm SE}^{0.5}$~\Mo. 
Similarly, from the size of the NE ellipse of 
$\timeform{1.9'} \times \timeform{0.9'}$, 
the plasma densities and the mass are respectively estimated to be 
$n_p \simeq n_e = 0.12~f_{\rm NE}^{-0.5}$~cm$^{-3}$ and 
$M = 57~f_{\rm NE}^{0.5}$~\Mo, where $f_{\rm NE}$ is the filling 
factor for the VAPEC component in the NE region.

Both the SE and NE plasmas were found to be in CIE condition. 
For full ionization equilibrium, the ionization timescale defined as 
$n_et$ is required to be $\geq~10^{12}$~cm$^{-3}$~s, 
where $t$ is the plasma age or the time since the gas was 
shock-heated (Masai 1984). Therefore, the plasma ages of the 
SE and NE plasmas are estimated to be higher than 
$\sim 9\times 10^4$~yr and $\sim 3\times 10^5$~yr, respectively. 
As mentioned in subsection~\ref{ssec:non-thermal}, the real age 
of the SNR is likely to be lower than $\sim 2\times 10^4$~yr. 
Thus, the plasma ages are significantly older than the SNR age. 
This fact can be interpreted that the SB shell had been ionized 
by preceding stellar winds and/or SNe, before the re-heating by 
the interior SNR. In CIE plasma, energy equipartition between 
electron and protons can be assumed. Therefore, the thermal 
energies ($E = 3n_e V kT_e$) of the SE and NE plasmas are estimated 
to be $6.7\times 10^{50}$~ergs and $1.2\times 10^{50}$~ergs,
respectively.

The electron temperature derived for the SE rim (0.58--0.76~keV) 
is significantly higher than the value of the previous report of B04 
(0.19--0.23~keV). If we fitted the SE spectrum fixing the temperature 
to 0.21~keV, the best-fit value of B04, the model failed to reproduce 
the fluxes of the Ne\emissiontype{X} and Mg\emissiontype{XII} K$\alpha$ 
lines, which are clearly detected in our spectrum (see figure~\ref{fig:se}). 
Indeed, the K$\alpha$-line flux ratios of 
Ne\emissiontype{X}/Ne\emissiontype{IX} and 
Mg\emissiontype{XII}/Mg\emissiontype{XI} predicted by the 0.21~keV 
CIE plasma are only less than 10\%. Since the emission lines from these 
hydrogen-like ions were not detected in the Chandra spectrum probably 
due to the poor photon statistics and relatively poor energy resolution, 
the systematically lower temperature would be obtained by B04.

\subsection{Non-Thermal Emission}
\label{ssec:non-thermal}

The high quality spectrum of the West rim rejected the simple 
power-law model. We found that either the broken power-law or the 
SRCUT was required to fit the West spectrum, for the first time.
Similar spectral steepening had been found in the synchrotron X-rays 
from the other SNRs, SN~1006 (Bamba et al.~2008) and  RX~J1713.7--3946 
(Takahashi et al.~2008). Therefore, the previous argument (B04) becomes 
more robust that the non-thermal X-rays from 30~Dor~C are synchrotron 
emissions from high energy electrons accelerated by the shock front 
of the SNR. 

In the SRCUT model for the West rim, the roll-off frequency was 
obtained to be (3--7)$\times 10^{17}$~Hz for the typical range of 
the spectral index at the radio band of $\alpha =$ 0.5--0.6. 
We also tried to fit the SE and NE spectra using the SRCUT models 
with $\alpha = 0.5$ instead of the power-law components. Then, the 
roll-off frequencies were obtained to be $> 1.1\times 10^{17}$~Hz 
and (0.7--1.4)$\times 10^{17}$~Hz, respectively. 
All these values are significantly higher than that of SN~1006 
($\sim 6\times 10^{16}$~Hz: Bamba et al.~2008), and similar to 
RX~J1713.7--3946 ($\sim 2\times 10^{17}$~Hz: Takahashi et al.~2008), 
the highest value among the shell-like SNRs. 
Since $\nu _{\rm rolloff} \propto BE_e^2$, where $B$ and $E_e$ are 
the magnetic field and the maximum energy of accelerated electrons 
(Reynolds 1998), the electrons would be accelerated to higher energy 
than that in the case of SN~1006, if the comparable strengths of the 
magnetic field are assumed.

30~Dor~C has a radius of $\sim 44$~pc, which is much larger than 
any other shell-like SNRs with non-thermal emission (e.g., 
RX~J1713.7--3946, RX~J0852.0--4622, RCW~86: $r < 20$~pc). 
If we simply assume that the blast-wave have been expanded almost freely 
in the cavity with the initial speed (possibly $\sim 10^4$~km~s$^{-1}$), 
then the SNR age would be $\sim 4\times 10^3$~yr. These may be the lower 
limit of the age. 
On the other hand, in order to emit intense synchrotron X-rays, 
the shock speed higher than $\sim 2000$~km~s$^{-1}$ is required 
by the standard diffusive shock acceleration theory 
(e.g., Aharonian \& Atoyan 1999). Therefore, the upper limit of 
the age is estimated to be $\sim 2\times 10^4$~yr. 
We thus conclude that the age of 30~Dor~C would be 
between $\sim 4\times 10^3$~yr and $\sim 2\times 10^4$~yr; 
30~Dor~C is the oldest among the known non-thermal SNRs.

Our observation demonstrates even older-system can be a site of 
higher energy acceleration than typical young SNRs, such as SN~1006. 
RCW~86 is another shell-like SNR with non-thermal shell, 
which has a relatively old age of $\sim 1800$~yr. 
Similarly to 30~Dor~C, RCW~86 also coincides with 
the OB association, and is considered to be fast expanding 
inside the large cavity (Vink et al.~2006; Yamaguchi et al.~2008). 
The long timescales of the efficient acceleration, achieved in 
30~Dor~C and RCW~86, would be due to the extremely low density of 
the cavity created inside the SB shell; 
the SNRs in SBs can be better cosmic-ray accelerator than 
the isolated SNRs in homogeneous ISM.

Non-thermal X-rays had been detected from other star-forming regions, 
RCW~38 (Wolk et al.~2002) and Westerlund~1 (Muno et al.~2006) in our 
Galaxy, and DEM~L192 (N51D: Cooper et al.~2004) in the LMC. 
However, the emissions from these regions show center-filled 
morphologies, in contrast to those from 30~Dor~C and RCW~86.
Moreover, their photon indexes (e.g., $\Gamma \sim 1.3$ for DEM~L192) 
are lower (harder) than those of typical shell-like SNRs. 
Therefore, the origins of these emissions are still unclear. 
Future detailed studies of the thermal and non-thermal properties of 
these star-forming regions using sensitive observatory, such as Suzaku, 
would help to reveal their origins. Observations of more SBs are also 
necessary to determine if non-thermal emissions are commonly present 
or not in SB regions.

\section{Summary}
\label{sec:summary}

We have analyzed Suzaku/XIS data of 30~Dor~C.
The results are summarized as follows:
 
\begin{enumerate}

\item  The West rim is dominated by the non-thermal emission, whereas 
  the SE and NE rims show the thermal and non-thermal emissions. 

\item  The spectrum from the West rim is well-reproduced by a power-law 
  with spectral cutoff. The cutoff frequency of $(3-7)\times 10^{17}$~Hz 
  is the highest, although the age is the oldest, among the known 
  shell-like SNRs with non-thermal emission. 

\item  The thermal components of the SE and NE spectra are 
  well-represented by the plasma at a collisional ionization equilibrium. 

\item  The electron temperature of the thermal plasma in the SE rim 
  ($kT_e \sim 0.7$~keV) is found to be higher than the value of 
  the previous report. 

\end{enumerate}

\bigskip
The authors would like to thank the Suzaku Science Working Group and 
the members of Suzaku SN~1987A team. We also thank Yasunobu Uchiyama 
for his useful comments. H.Y.~is supported by the Special Postdoctoral 
Researchers Program in RIKEN. A.B.~is supported by JSPS Research 
Fellowship for Young Scientists.


\end{document}